\documentclass[a4paper]{jpconf}
\usepackage{graphicx}
\usepackage{cite}
\usepackage[intlimits]{amsmath}
\usepackage{amssymb}
\usepackage{gensymb}
\usepackage{multido}

\newcommand{\rr}{\textbf{r}}

\begin{document}
\title{Spectrum and fraction of cosmic ray positrons: results of the anomalous diffusion approach}

\author{Nikolay Volkov, Anatoly Lagutin, Alexander Tyumentsev}

\address{Altai State University, Radiophysics and Theoretical Physics Department, Barnaul, Russia, 656049, Lenin ave., 61}

\ead{volkov@theory.asu.ru, lagutin@theory.asu.ru, tyumentsev@theory.asu.ru}

\begin{abstract}
We present the results of new calculations of the energy spectra of cosmic ray electrons, positrons and also positron fraction under assumption that both electrons and positrons are generated by the same Galactic sources, which accelerate particles with same power-law injection spectral index $p$. The value of the injection index $p\approx 2.85$ retrieved in our works from the analysis of the observed CR spectra has been used. The propagation of particles through the highly non-homogeneous  interstellar medium has been described by the anomalous diffusion model.

We show that proposed approach allows the self-consistent description of the recent AMS-02 data. In contrast to the conclusion made by AMS-02 collaboration about the different origin of the high-energy positrons and electrons, our results demonstrate that the differing behavior of spectral indices of electrons and positrons with energy can be described under the assumption that both positrons and electrons have common Galactic sources.

Our prediction (Lagutin et al. Bull. Russian Academy of Sciences. Physics, 2013, Vol. 77, No. 11, p. 1312) that the  $e^{+}$  to total $e^{-} + e^{+}$ ratio reaches a constant value for E $\gg 200$ GeV is confirmed. 
\end{abstract}

\section{Introduction}

According to standard cosmic ray (CR)  electron/positron  origin scenario~\cite{Berezinsky:1990,Strong:2007,Grasso:2011}, electron sources are continuously distributed in Galactic disk and positrons are produced only by collisions of primary CR nuclides with Galactic interstellar medium. In this CR origin picture the ratio of the secondary $e^{+}$  to total $e^{-} + e^{+}$ flux decreases with increasing energy. However, new experimental results obtained in the last decade by PAMELA~\cite{pamela:el}, Fermi-LAT~\cite{fermi} and AMS-02~\cite{ams02:elpos} collaborations demonstrate steadily increasing of the positron fraction (PF) from 10 GeV to $\sim$ 300 GeV.  This is not consistent with only secondary mechanism of positron production and probably with the standard paradigm of the Galactic CR origin. The presence of new primary positron component seems unavoidable. 

Models, ranging from the annihilation/decay of dark matter (DM) to the new astrophysical sources, have been proposed in order to explain the rise in the PF  (see a review~\cite{Serpico:2012} and the references therein; see also~\cite{Blasi:2009,Kachelrie:2011,Berezhko:2013,Erlykin:2013,Gaggero:2013,
Cholis:2013,DiMauro:2014,Cowsik:2014,Mertsch:2014}). Leading proposals put forth to explain reported PF results involve $e^{\pm}$ pair emission by nearby pulsars. These new charge symmetric sources of electrons and positrons are introduced in order to account for the "excess" of the positrons relative to the background fluxes, associated to galactic CR sources of the standard electron/positron  origin scenario.
 
In this paper, the so-called alternative explanation (see~\cite{Blasi:2009,Kachelrie:2011,Berezhko:2013,Erlykin:2013}) of the rise in the PF which involves the positrons production in the astrophysical sources of the CR is investigated. We assume that both positrons and electrons have the common Galactic sources accelerated the particles with the same power-law injection spectral index $p$ (see~\cite{Kachelrie:2011}). The value of the injection index $p\approx 2.85$ retrieved in our works~\cite{Lagutin:2001np,Lagutin:2004a} from analysis of the observed CR spectra is used. We also suppose that the propagation of particles through the highly non-homogeneous  interstellar medium is described by the anomalous diffusion model~\cite{Lagutin:2001np,Lagutin:2003,Lagutin:2004a}. The anomaly in this model results from large free paths (L\'{e}vy flights) of particles between galactic inhomogeneities in the fractal-like Galactic medium. First results related to the PF behavior in this approach were discussed in our papers~\cite{Volkov:2011,Volkov:2013}.

The main goal of this paper is to validate the key assumptions of proposed approach using new results of the AMS-02 collaboration.

\section{Anomalous diffusion model}

In the standard scenario, the propagation of galactic cosmic ray electrons and positrons through the
interstellar medium (ISM) is described by Syrovatskii's diffusion equation for density of particles 
$N(\rr,t,E)$  of energy $E$~\cite{Sirovat:1959} 
\begin{equation}\label{Syr}
\frac{\partial N(\rr,t,E)}{\partial t}=D(E)\triangle N(\rr,t,E)+\frac{\partial(b(E)N(\rr,t,E))}{\partial E} + S(\rr,t,E).
\end{equation}
In this equation $D(E)$ is the diffusion coefficient, $b(E)$ -- the rate of change of the particles energy during their propagation through the interstellar medium and $S(\rr,t,E)$ -- the source function of electrons/positrons.

This equation determines propagation process of particles (electrons/positrons) under the approximation that magnetic field inhomogeneities of the ISM are small-scale and can be considered as a homogeneous Poisson ensemble. However, during a few last decades many evidences of the existence of multiscale structures in the Galaxy have been found. Filaments, shells, clouds are entities widely spread in the ISM. A rich variety of structures that are created in the ISM can be related to the fundamental property of turbulence called intermittency. In such a fractal-like ISM we certainly do not expect that Syrovatskii's diffusion equation remains accurate enough.

Generalization of the equation~\eqref{Syr} leads to anomalous diffusion model~\cite{Lagutin:2001np,Lagutin:2003,Lagutin:2004a}. In this paper, instead of Syrovatskii's diffusion equation, the superdiffusion equation, proposed in our works~\cite{Volkov:2009,Volkov:2011,Volkov:2013} is used:
\begin{equation}\label{SuperEq}
\frac{\partial N(\rr,t,E)}{\partial t}=-D(E,\alpha)(-\triangle)^{\alpha/2} N(\rr,t,E)+\frac{\partial(b(E)N(\rr,t,E))}{\partial E} + S(\rr,t,E).
\end{equation}
Here $D(E,\alpha)=D_0(\alpha)E^{\delta}$ is the anomalous diffusivity. Energy losses  $b(E)$ are presented in the form 
\begin{equation}\label{eq:losses}
b(E) = b_0 + b_1 E + b_2 E^2 \approx b_2 (E+E_1)(E+E_2),
\end{equation}
where $b_0 = 3.06\cdot 10^{-16}n$ (GeV s$^{-1}$) is rate of the ionization losses, $b_1 = 10^{-15}n$ (s$^{-1}$) determines the bremsstrahlung rate, and $b_2 = 1.38\cdot 10^{-16}$ (GeV$\cdot$ s)$^{-1}$ is the factor to take into account the inverse Compton losses and synchrotron radiation (for the intensity of the interstellar magnetic field $B = 5$~$\mu$G and background photon density $\omega = 1$~eV/cm$^{3}$). In~\eqref{eq:losses} we have introduced the notation $E_1 = b_0/b_1$ and $E_2 = b_1/b_2$.

The fractional Laplacian (Riesz operator) $(-\triangle)^{\alpha/2}$~\cite{Samko:1993} describes the large free paths (L\'{e}vy flights) of particles between Galactic inhomogeneities in the fractal-like Galactic medium.

We note that in case of $\alpha=2$ equation~\eqref{SuperEq} coincides with Syrovatskii's equation~\eqref{Syr}.

Green's function $G(\rr,t,E;E_0)$, corresponding to~\eqref{SuperEq},  can be found from an equation  
\begin{multline}\label{GreenEq_el}
\frac{\partial G(\rr,t,E;E_0)}{\partial t}=-D(E,\alpha)(-\Delta)^{\alpha/2} G(\rr,t,E;E_0)+\\+\frac{\partial b(E)G(\rr,t,E;E_0)}{\partial E}+\delta({\rr})\delta(t)\delta(E-E_0).
\end{multline}

\begin{figure}[!hb]
\includegraphics[width=\textwidth]{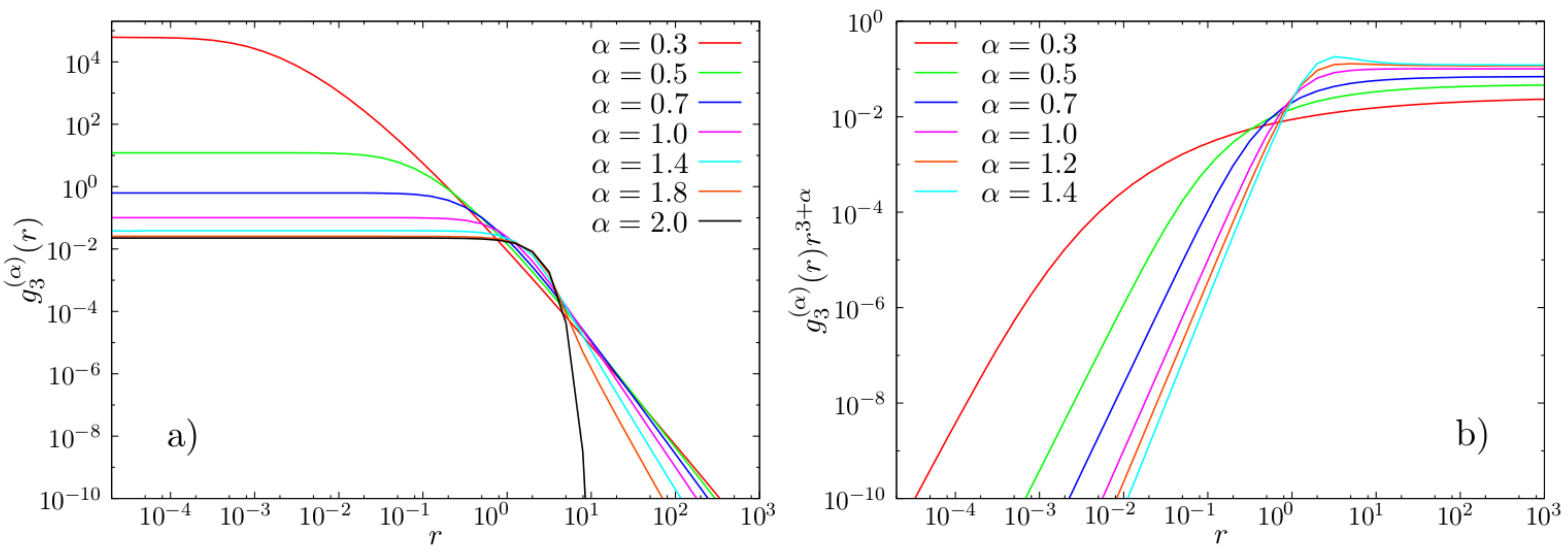}
  \caption{a) the probability density of three-dimentional sphericaly-symmetrical stable distribution $g_3^{(\alpha)}(r)$, b) the asymptotic behavior $r^{-3-\alpha}$ for $r\rightarrow \infty$}\label{fig:g3}
\end{figure}

With the use of the standard Syrovatskii substitutions~\cite{Sirovat:1959}
\begin{equation*}\label{eq:Sirovatsubs}
t' = t - \tau,\quad \tau(E,E_0)=\int\limits_E^{E_0}\frac{dE'}{b(E')},\quad
\lambda(E,E_0)=\int\limits_E^{E_0}\frac{D(E',\alpha)}{b(E')}dE',
\end{equation*}
and Fourier transform, it is easy to find 
\begin{eqnarray}\label{green}
G(\rr,t,E;E_0) =\frac{g_3^{(\alpha)}(|\rr|\lambda^{-1/\alpha})}
{\lambda^{3/\alpha} (1-b_2 t (E+E_2))^2}\delta\left(E_0-\left\{\frac{E+E_1}{1-b_1 t(E+E_2)/(E_2-E_1)}-E_1\right\}\right)\times\nonumber\\
\times H(1-b_2 t (E+E_2))H(t).
\end{eqnarray} 
Here
\begin{eqnarray}
E_0(t) = \frac{E+E_1}{1-b_1 t (E+E_2)/(E_2-E_1)}-E_1\nonumber
\end{eqnarray}
and $g_3^{(\alpha)}(r)$ is the probability density of three-dimentional sphericaly-symmetrical stable distribution~\cite{Uchaikin:1999}. The probability density of $g_3^{(\alpha)}(r)$ and the asymptotic behavior for $r\rightarrow \infty$ is shown in figure~\ref{fig:g3}.

\section{Solution of the anomalous diffusion equation}

Using the Green's function~\eqref{green}, we can write a solution of the superdiffusion equation~\eqref{SuperEq} for particles generated by sources, defined by the source function $S(\rr,t,E)$
\begin{equation*}
 N(\rr,t,E) = \int\limits_{\mathrm{R}^3} d \rr_0 \int\limits_E^{\infty} d E_0 \int\limits_{-\infty}^t d t_0
G(\rr,t,E;\rr_0,t_0,E_0)S(\rr_0,t_0,E_0).
\end{equation*}

\subsection{Point pulse source}

For a point pulse source with a power-law energy spectrum $S(\rr,t,E)=S_iE^{-p}\delta(\rr)\Theta(T-t)\Theta(t)$, which simulates generation of particles in supernovae, the following solution takes place.
\begin{multline}\label{eq:local}
N(\rr,t,E)=S_i \int\limits_{\mathrm{max}[0,t-T]}^{\mathrm{min}[t,1/b_2(E+E_2)]}dt' E_0(t')^{-p} \lambda(t',E)^{-3/\alpha}\times\\
\times (1-b_2 t'(E+E_2))^{-2} g_3^{(\alpha)}\left(|\rr|\lambda(t',E)^{-1/\alpha}\right).
\end{multline}

\subsection{Point steady source}

For a point steady source $S(\rr,E)=S_c E^{-p}\delta(\rr)$ we find
\begin{multline}\label{eq:global}
N(\rr,E)= S_c \int\limits_0^{1/b_2(E+E_2)}dt'E_0(t')^{-p}\lambda(t',E)^{-3/\alpha}\times\\
\times(1-b_2 t' (E+E_2))^{-2} g_3^{(\alpha)}\left(|\rr|\lambda(t',E)^{-1/\alpha}\right).
\end{multline}

\section{Energy spectrum of electrons and positrons}

The particle intensity $J^{\pm}(\rr,t,E) = (v/{4\pi})N^{\pm}(\rr,t,E)$  was expressed as
\begin{equation}\label{spectr_el}
J^{\pm}(\rr,t,E) = J^{\pm}_{\text{\it \tiny L}}(\rr,t,E) + J^{\pm}_{\text{\it \tiny G}}(\rr,E) + J^{\pm}_{\text{sec}}
\end{equation}
where $J^{\pm}_{\text{\it \tiny L}}$ is the local component, i.e. the contribution from nearby ($r< 1$ kpc) young ($t< 10^6$~yr) sources,  $J^{\pm}_{\text{\it \tiny G}}$ is the global spectrum component determined by the multiple old ($t\geq 10^6$ yr) distant ($r\geq 1$ kpc) sources and $J^{\pm}_{\text{sec}}$ is the secondary electron and positron fluxes originated
from interactions of primary cosmic rays with the interstellar medium.

To describe the distribution of nearby young sources simulation of the Poisson ensemble of sources was carried out. The Poisson distribution parameter (average number of sources in the local region) was chosen $\sim10$. This estimation corresponds to number of the well-known nearest supernova remnants and pulsars with $t< 10^6$~yr~\cite{Hooper:2009,Gendelev:2010}. Coordinates and times of birth of the sources were generated randomly and uniformly in the space region $r<10^3$~pc and in the time interval $10^4\leq t< 10^6$~yr. The duration of particle generation by local sources was assumed to be $T\approx10^4$~yr.

Distribution of multiple old ($t\geq 10^6$ yr) distant ($r\geq 1$ kpc) sources was considered according to standard leptonic scenario (i.e. as the system of steady-state sources). To accommodate the contribution of secondary electrons and  positrons produced in the collision of cosmic rays nuclei with the interstellar medium, the GALPROP code~\cite{Moskalenko:1998} was used with the parameters stated in our model.

At energies $E\lesssim 10$~GeV, the fluxes of both electrons and positrons observed in the Solar system are influenced by modulation effects. To take into account the solar modulation we used the model~\cite{Axford:1968} with the potential $\Phi = 600$~MV.

\section{Estimation of the model parameters}

Estimations of key model parameters (injection spectral index $p$ and exponent $\delta$ describing the energy dependence of the anomalous diffusivity) were obtained in~\cite{Lagutin:2004a} from the analysis of observed CR spectra. It was shown in~\cite{Lagutin:2004a}  that with the use of representation $N=N_0E^{-\eta}$ for observed CR spectra and the values of the exponents below and above the knee energy 
$E_{\text{knee}}$, 
$$\eta|_{E \ll E_{\text{knee}}} = p - \delta,$$
$$ \eta|_{E \gg E_{\text{knee}}} = p + \delta,$$
the variation of spectral exponent $\Delta\eta = \eta|_{E \gg E_{\text{knee}}} - \eta|_{E \ll E_{\text{knee}}}$  can be found as $\Delta\eta = 2\delta.$
Since $\eta|_{E \ll E_{\text{knee}}}\sim 2.62$ and $\eta|_{E \gg E_{\text{knee}}}\sim 3.16$, from the  equalities above it is easy to retrieve the injection spectral index $p$ and exponent $\delta$ of anomalous diffusivity:    
$$ p\approx 2.85, \quad \delta\approx 0.27.$$

We note that the estimation of $\delta\approx 0.27$ almost coincides with the value of $1/3$, accepted for the Kolmogorov turbulence~\cite{Berezinsky:1990}. Adopted value $p\approx 2.85$ for the CR sources is supported by the recent results on SNRs: for the SNR RX J1713.7-3946 the exponent $p\approx 3.0$ was obtained~\cite{Tanaka:2008}; the Fermi-LAT experiment reported $p\approx 2.87$ for $E>69$~GeV for the SNR IC 443~\cite{Abdo:2010ic443,Tavani:2010} and $p\approx 3.3$ for the SNR W44~\cite{Abdo:2010w44}.

To assess parameter $\alpha$, which is specific for our model, the results of studies of particles diffusion in cosmic and laboratory plasma have been used. It is known that for the superdiffusion regime of propagation ($1<\alpha <2$) the diffusion packet spreads with time as $\Delta x^2 \sim t^{2/\alpha}$. Interpretation of the data for magnetosphere leads to conclusion that the width of the diffusion packet increases with the rate $\Delta x^2 \sim t^{1.4}$~\cite{Greco:2003}. Investigation of the particles propagation before the shock wave in the circumterrestrial plasma gives the dependence of $\Delta x^2\sim t^{1.19}$~\cite{Perri:2008}. According to these results we obtain the estimation as $\alpha \approx 1.4$.

Finally, for the anomalous diffusivity $D(E,\alpha) = D_0(\alpha)E^{\delta}$ we get the value of $D_0(\alpha)\approx 2 \cdot 10^{-4}$~pc$^{1.4}$/year~\cite{Lagutin:2004a}.

\begin{figure}[!hp]
\begin{center}
\includegraphics[width=.85\textwidth]{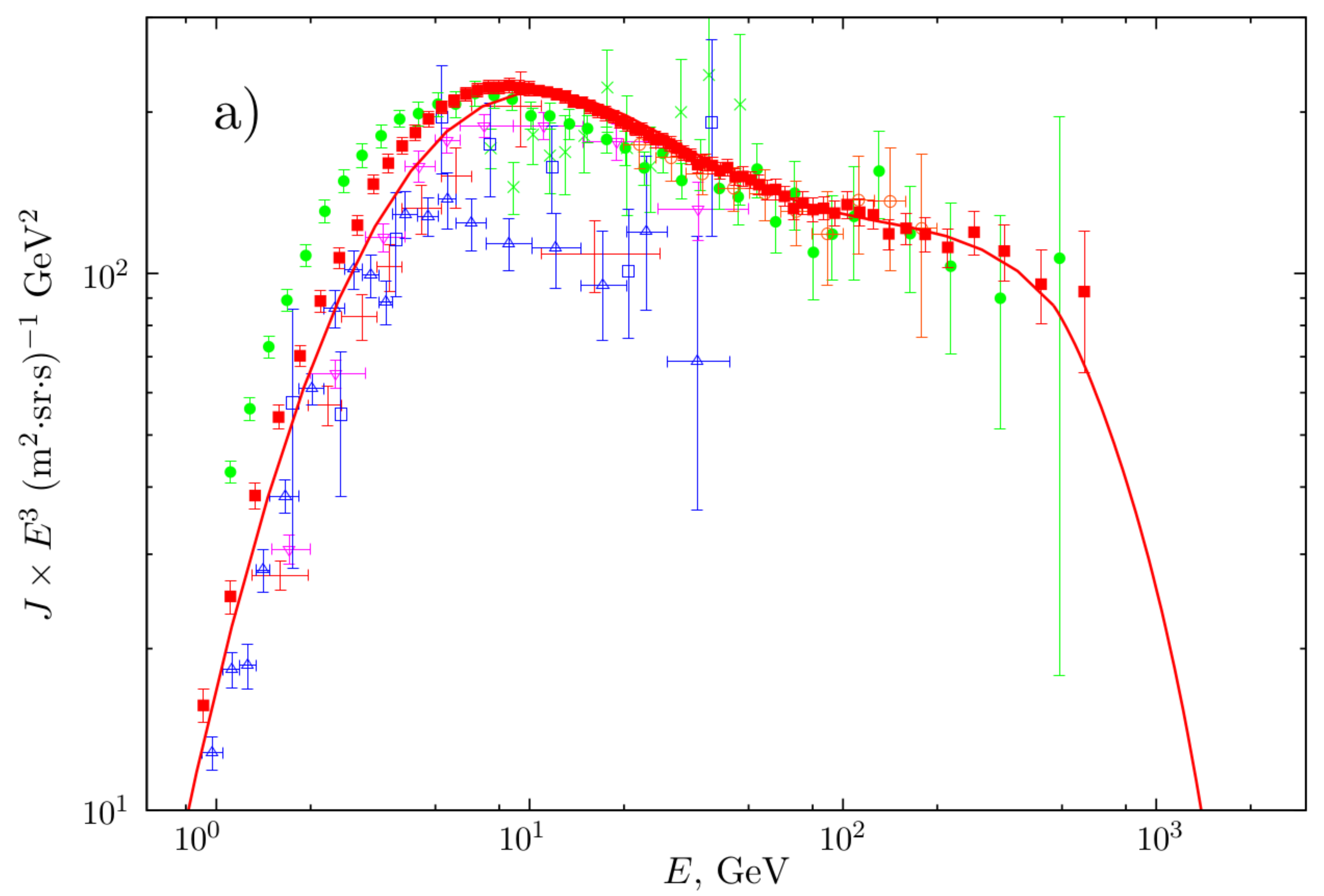}
\includegraphics[width=.85\textwidth]{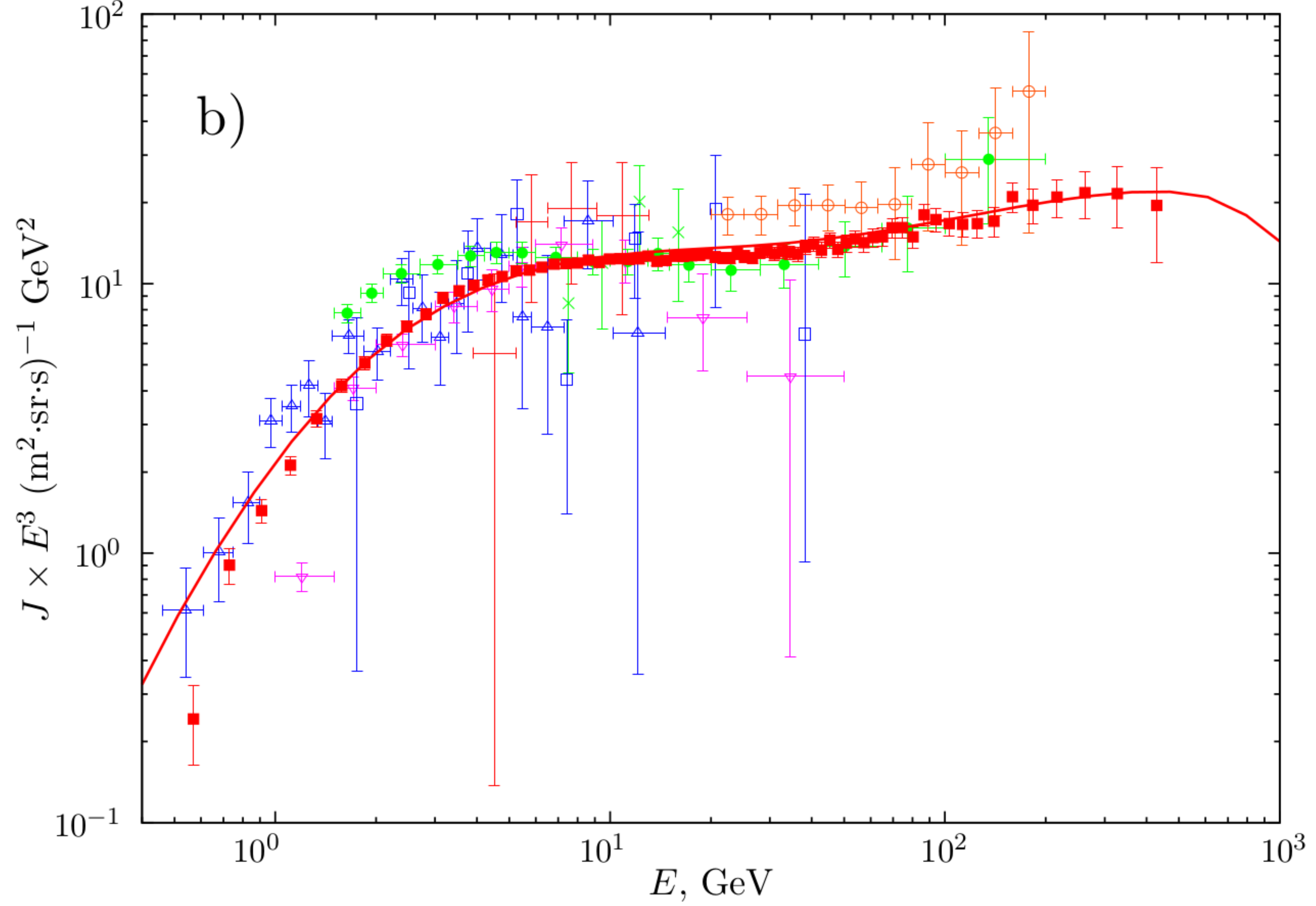}
\includegraphics[width=.85\textwidth]{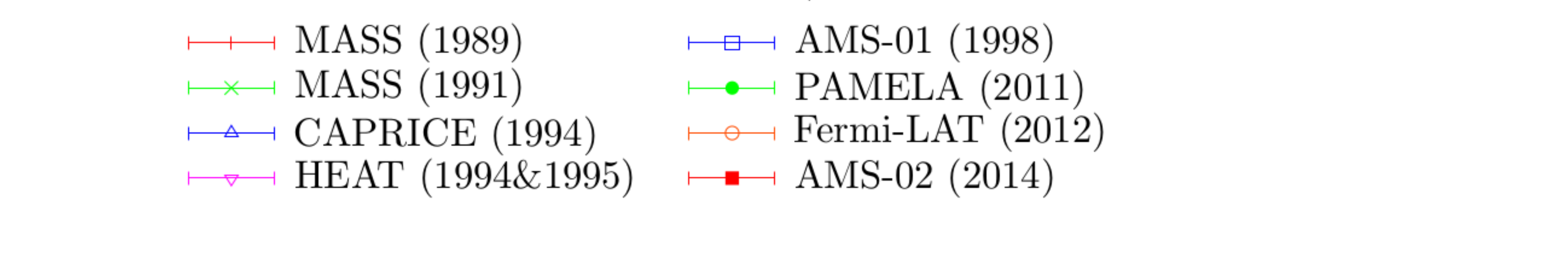}
\end{center}

\caption{Spectrum (times $E^3$) of electrons a) and positrons b) in the proposed model. Experimental data from~\cite{mass:1989,mass:1991,caprice:1994,heat9495,ams01elpos,pamela:el,fermi,ams02:elpos} }\label{fig:elposspectr}
\end{figure}

\begin{figure}[t]
\begin{center}
\includegraphics[width=.9\textwidth]{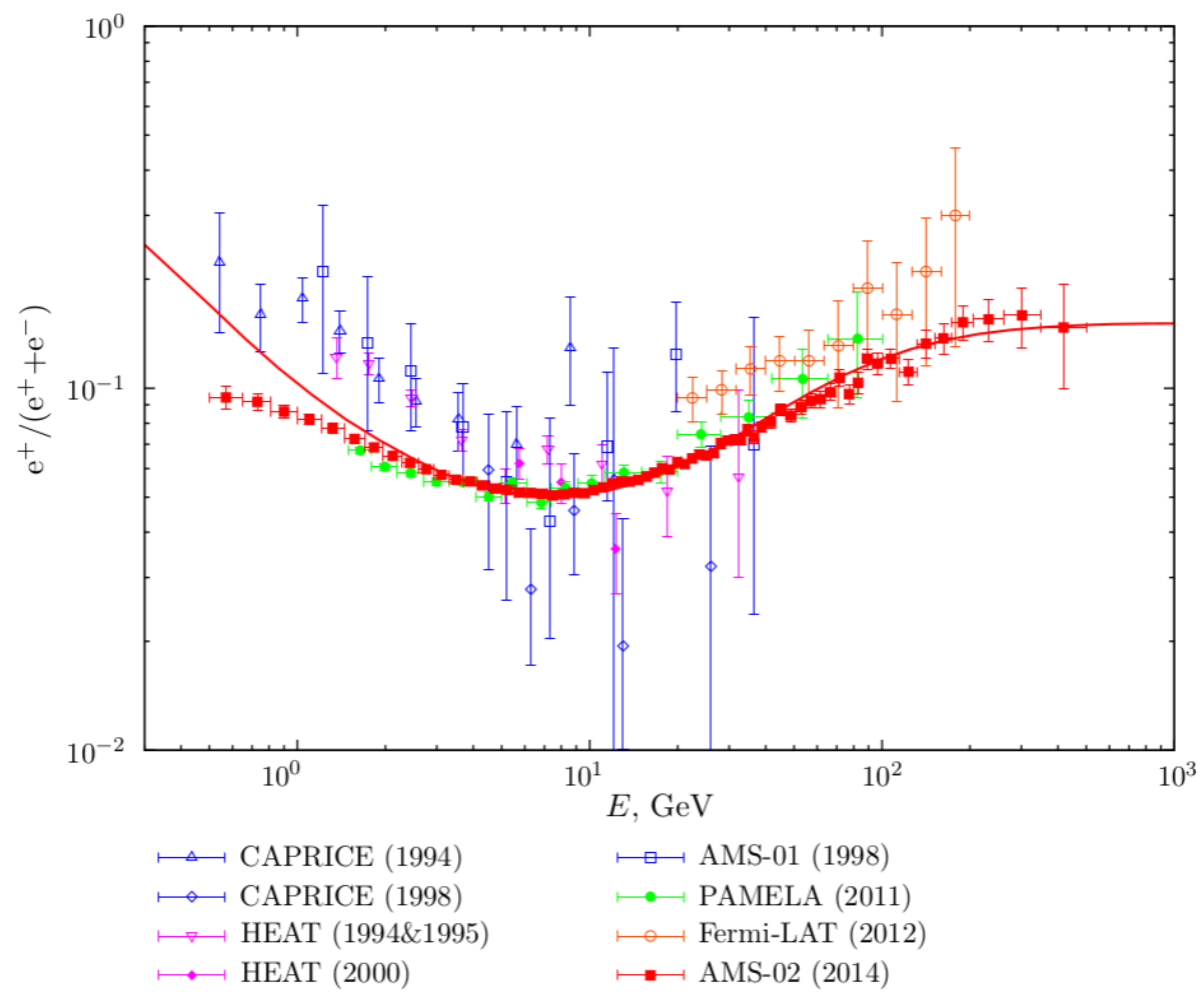}
\end{center}

\caption{Positrons fraction in the proposed model. Experimental data from~\cite{mass:1989,mass:1991,caprice:1994,heat9495,ams01elpos,pamela:el,fermi,ams02:elpos,caprice:1998,
heat:2000,pamela:pos,ams02:frac}}\label{fig:posfract} 
\end{figure}

Results of our calculations of the energy spectra of cosmic ray electrons and positrons as well as positron fraction obtained in the framework of the proposed approach are shown in figures~\ref{fig:elposspectr}--\ref{fig:elposspectrall}.

\section{Results and conclusions}

We present the results of new calculations of the energy spectra of cosmic ray electrons, positrons and also positron fraction under assumption that both positrons and electrons are generated by the same Galactic sources, which accelerate particles with same power-law injection spectral index $p$. The value of the injection index $p\approx 2.85$ retrieved in our works~\cite{Lagutin:2001np,Lagutin:2004a} from the analysis of the observed CR spectra has been used. The propagation of particles through the highly non-homogeneous  interstellar medium has been described by the anomalous diffusion model~\cite{Lagutin:2001np,Lagutin:2003,Lagutin:2004a}. The anomaly in this model results from large free paths (L\'{e}vy flights) of particles between galactic inhomogeneities in a fractal-like Galactic medium.

\begin{figure}[h]
\includegraphics[width=\textwidth]{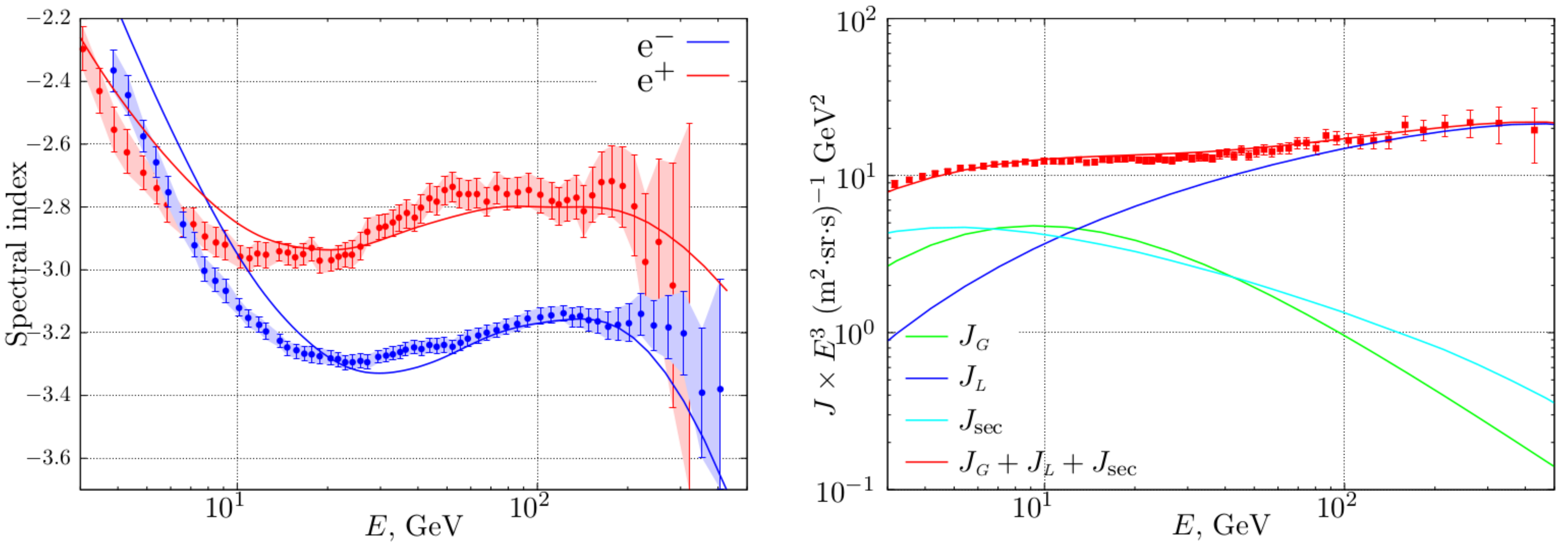}
\caption{Spectral indices of the fluxes of electrons and positrons in the framework of the proposed model (color lines). Points are the results of AMS-02 collaboration~\cite{ams02:elpos}}\label{fig:elposspectralindex}
\caption{Positrons spectrum (times $E^3$) in the proposed model. Contribution from global sources (green), local group sources (blue) and secondary positrons (cyan) also presented}\label{fig:elposspectrall}
\end{figure}

Figures~\ref{fig:elposspectr}--\ref{fig:elposspectrall} clearly show that proposed approach allows the self-consistent description of the recent AMS-02 data on electrons, positrons and positron fraction.

In contrast to the conclusion made by AMS-02 collaboration in~\cite{ams02:elpos} about the different origin of the high-energy positrons and electrons, our results demonstrate that the differing behavior of spectral indices of electrons and positrons with energy can be described under the assumption that both positrons and electrons have common Galactic sources. We found that the difference in spectral indices is due to the fact that in the energy region between 20 and 200~GeV in addition to the yield from the primary positron sources there is also a significant contribution of secondary positrons (see figure~\ref{fig:elposspectrall}).

The positron fraction obtained in this paper (figure~\ref{fig:posfract}) confirms our prediction~\cite{Lagutin:2013} that the  $e^{+}$  to total $e^{-} + e^{+}$ ratio reaches a constant value for E $\gg 200$ GeV.

\section*{Acknowledgement}
This work was supported in part by the Russian Foundation for Basic Research, project no.~14-02-31524.
We also acknowledge helpful comments and suggestions from an anonymous referee.

\section*{References}

\bibliographystyle{iopart-num}
\bibliography{ecrs}

\end{document}